\documentclass[showpacs,aps,graphicx,twocolumn]{revtex4}
\usepackage{stmaryrd}
\usepackage{txfonts}
\usepackage{mathrsfs}
\usepackage{amsfonts}
\usepackage{amssymb}
\usepackage{graphicx}
\usepackage{eufrak}
\usepackage{multirow}
\usepackage{bm}

\newcommand{\tabincell}[2]{\begin{tabular}{@{}#1@{}}#2\end{tabular}}

\begin{document}


\title{Optimal synthesis of multivalued quantum circuit}

\author{Yao-Min Di$^{1}$\footnote{Corresponding author:
yaomindi@sina.com}, Hai-Rui Wei$^2$}
\address{$^{1}$School of Physics $\&$ Electronic Engineering, Jiangsu Normal
University, Xuzhou 221116,  China
\\$^{2}$Department of Mathematics and Mechanics, School of Applied Science,
University of Science and Technology Beijing, Beijing 100083, China}

\date{\today }

\begin{abstract}

Although many of works have been done in multivalued quantum logic synthesis, the question whether multivalued quantum circuits are more efficient than the conventional binary quantum circuits is still open. In this article we devote to the optimization of generic multivalued quantum circuits. The multivalued quantum Shannon decompositions (QSD) are improved so that the circuits obtained are asymptotically optimal for all dimensionality $d$. The syntheses of uniformly multifold controlled $R_y$ rotations are also optimized to make the circuits further simplified. Moreover, the theoretical lower bound of complexity for multivalued quantum circuits is investigated, and a quantity known as efficiency index is proposed to evaluate the efficiency of synthesis of various quantum circuits. The algorithm for qudit circuits given here is an efficient synthesis routine which produces best known results for all dimensionality $d$, and for both cases the number of qudit $n$ is small and that is asymptotic. The multivalued quantum circuits are indeed more efficient than the binary quantum circuits. The facts, the leading factor of the lower bound of complexity for qudit circuits is small by a factor of $d-1$ in comparison to that for qubit circuits and the asymptotic efficiency index is increased with the increase of dimensionality $d$, reveal the potential advantage of qudit circuits over generic qubit circuits. The generic $n$-qudit circuits with $d\geq5$ and generic two-ququart circuits synthesized by the algorithm given here are practical circuits which are more efficient than the most efficient qubit circuits.

\end{abstract}


\maketitle \flushbottom

\section{Introduction} \label{sec1}

Enormous progress has been made in the field of quantum information science over the past two and a half decades. Most approaches to quantum information processing use two-level quantum systems (qubits). However there is increasing interest in exploiting protocol with multilevel quantum systems (qudits) \cite{1,2,3,4,5,6,7,8,9}. The simplest multilevel system, the three-level quantum system, is called a qutrit, the four-level quantum system is called a ququart. The multivalued quantum information is exciting because quantum systems usually have multi-levels, it enables us to full use of various resources.

 In quantum computing, the algorithms are commonly described by the quantum circuit model. The process of constructing quantum circuits by some elementary components is called synthesis. The complexity of quantum circuit can be measured in terms of the number of elementary gates required. A large amount of work in these areas has been done for binary quantum computing \cite{10,11,12,13,14,15,16,17,18,19,20,21}. The CNOT gate is one of most widely used two-qubit elementary gate. It has been shown that the CNOT gate with one-qubit gate is universal for qubit quantum circuits \cite{10,11}. The best result so far for the synthesis of generic qubit quantum circuits are given by Shende \emph{et al.} based on quantum Shannon decomposition (QSD) \cite{19}.

Although many of works also have been done in multivalued quantum logic synthesis \cite{22,23,24,25,26,27,28,29}, the works on this area are still far from complete. The results obtained cannot show the advantage of using multilevel quantum systems in the complexity of quantum logic synthesis. Which gate is chosen as the two-qudit elementary gate of the qudit quantum circuit is a crucial issue for multivalued quantum computing, and there have been many proposals. In our recent previous work, the generalized controlled $X$ (GCX) gate has been proposed as the two-qudit elementary gate for the multivalued circuits \cite{30,31}. We generalize QSD, the most powerful technique for the synthesis of generic qubit circuits, to the multivalued case. Based on the GCX gate, using the multivalued QSD, we obviously improve the results of the synthesis of qudit quantum circuit \cite{31}. But there are still some problems. One is that the quantum circuits built by multivalued QSD algorithm are not asymptotically optimal except for the dimensionality of qudit $d$ is a power of two. It was not clear whether we can build efficient quantum circuits for the qudit $d$ is not a power of two as that $d$ is a power of two. The other is that the multivalued quantum circuits in Ref. \cite{31} do not show obvious advantage over the circuits for binary systems. The problem whether the multivalued quantum circuits can be more efficient than the binary circuit is still open.

In this article, we devote to optimizing multivalued quantum circuits and to solving the problems stated above. The multivalued QSD for the qudit $d$ is not a power of two is optimized so that the synthesis of quantum circuits for these qudit also is asymptotically optimal. The synthesis of the uniformly multifold controlled  $R_y$ rotations is also optimized to make the circuits further simplified. The theoretical lower bound of complexity for qudit quantum circuits is investigated. A quantity known as efficiency index is proposed to evaluate the efficiency of synthesis of generic $n$-qudit circuits. The results and comparison show that algorithm given here is most efficient qudit synthesis routine so far which produces best known results in all respects. The multivalued quantum circuits are indeed more efficient than the binary quantum circuits.

The article is organized as follows: The lower bound of complexity for qudit circuits is investigated in Sec. \ref{sec-2}. The leading factor of the lower bound of complexity for qudit circuits is small by a factor of $d-1$ in comparison to that for qubit circuits. The optimization of the multivalued QSD and uniformly multifold controlled $R_y$ rotations, the structure and the GCX gate count of optimal qudit circuits are given in Sec. \ref{sec-3}. The efficiency of synthesis of quantum circuits is discussed in Sec. \ref{sec-4}. The quantity in term of efficiency index is proposed in this section. The asymptotic efficiency index is increased with the increase of dimensionality $d$ for these circuits. The efficiency indexes of generic $n$-qudit circuits with   $d\geq5$ and generic two-ququart circuits given here are higher than that of the most efficient qubit circuits. Finally, a brief conclusion and future work are given in Sec. \ref{sec-5}.

\section{Lower Bounds of GCX gates} \label{sec-2}

The GCX gate (denoted as $\text{GCX}(m\rightarrow X^{(ij)}$)) is a controlled-$U$ two-qudit gate which implements the  $X^{(ij)}$ operation on the target qudit iff the control qudit is in the state $|m\rangle$, where  $X^{(ij)}=|i\rangle \langle j| +|j\rangle\langle i|+\sum_{k\neq i,j} |k\rangle\langle k|$. The GCX gate essentially is a CNOT gate. For a multilevel quantum system which forms a qudit, two levels in the system forms a qubit. If a two-qubit CNOT gate is realized in two such systems, a GCX gate is naturally obtained. The number of the GCX gates required can be used as a unified measure for the complexity of various quantum circuits \cite{31}.

In quantum computing, the quantum circuit is a unitary transformation on the quantum states. A generic $n$-qudit quantum circuit is fully determined by $d^{2n}-1$ real parameters (up to a phase factor). Here the qudit circuits are constructed by using GCX gates and arbitrary one-qudit gates. The GCX gates do not introduce any parameters, but they are a kind of barriers that separate one-qudit gates so that they cannot merge into a resulting one-qudit gate for each qudit. Intuitively, every GCX gate can be accompanied two one-qudit gates, one is for control qudit $G_1$ and the other for target qudit $G_2$ , applied after every GCX gate. Two one-qudit gates can carry $2(d^2-1)$  real parameters, but only part of them can be carried by a GCX gate.

The one-qudit gate corresponds to a  $SU(d)$ group, and the $d^2-1$ parameters correspond to the bases of  $su(d)$ algebra. Without loss of generality, we consider the $\text{GCX}(d-1\rightarrow X^{(d-2,d-1)})$ gate and use the natural bases of $u(d)$ algebra $|i\rangle\langle j|$, where $i, j\in 0,1,\cdots,d-1$. There are $2(d-1)$ bases which do not commute with the GCX gate in $G_1$, they are $|d-1\rangle\langle i|$ and $|i\rangle\langle d-1|$, where $i\in 0,1,\cdots, d-2$, so the gate can separate $2(d-1)$  parameters in $G_1$. There are $4(d-1)$  such bases in $G_2$, they are $|d-1\rangle\langle i|$, $|d-2\rangle\langle i|$,  where $i\in 0, 1, \cdots, d-1$ and  $|j\rangle\langle d-1|$, $|j\rangle\langle d-2|$, where $i\in 0, 1, \cdots,d-3$. But there are $2(d-1)$  linear combinations of them which commute with the GCX gate, they are  $|d-2\rangle\langle i|+|d-1\rangle\langle i|$, where $i\in 0,1,\cdots, d-1$, and $|j\rangle\langle d-2|+|j\rangle\langle d-1|$, where $j\in 0,1,\cdots, d-3$. The GCX gate also can separate $2(d-1)$  independent parameters in $G_2$. Each GCX gate can bring at most $4(d-1)$ parameters. For a generic $n$-qudit quantum circuits, there are $d^{2n}-n(d^2-1)-1$ parameters which need to be brought by the GCX gates. The theoretical lower bound of complexity for generic qudit circuits is $(d^{2n}-n(d^2-1)-1)/(4(d-1))$. The lower bound of complexity for generic qubit circuits is $(4^n-3n-1)/4$ \cite{16}. The leading factor of the lower bound of complexity for qudit circuits is small by a factor of $d-1$ in comparison to that for qubit circuits reveals the potential advantage of qudit circuits over the qubit circuits.

\section{Optimal Synthesis of Multivalued Quantum Circuits} \label{sec-3}

\subsection{Synthesis Based on Multivalued QSD \cite{31}} \label{sec-3-1}

The first phase of multivalued QSD is to use the cosine-sine decomposition (CSD) \cite{32}. Let the $m\times m$ unitary matrix $W$ be partitioned in $2\times2$ block form as
\begin{equation}   \label{eq-1}
W=\bordermatrix {  &r      & m-r    \cr
                 r &W_{11} & W_{12} \cr
               m-r &W_{21} & W_{22} \cr},
\end{equation}
with $2r\leq m$. Here $r$ is called as partition size. Let $W=U\Gamma V$ be CSD of the matrix, then
\begin{equation}   \label{eq-2}
U=\bordermatrix {  &r     & m-r    \cr
                 r &U_{1} & 0 \cr
               m-r &0     & U_{2} \cr},
\end{equation}
\begin{equation}   \label{eq-3}
\Gamma=\bordermatrix {  &r & r    & m-2r    \cr
                 r &C & -S & 0    \cr
                 r &S & C  & 0    \cr
               m-2r &0 & 0  & I    \cr},
\end{equation}%
\begin{equation}   \label{eq-4}
V=\bordermatrix {  &r      & m-r    \cr
                 r &V_{1} & 0 \cr
               m-r &0     & V_{2} \cr},
\end{equation}
where $C$ and $S$ are diagonal matrices of the forms $C=diag\{\cos\theta_1,\cos\theta_2,\cdots,\cos\theta_r\}$ and $S=diag\{\sin\theta_1,\sin\theta_2,\cdots,\sin\theta_r\}$, $I$ is the $(m-2r)\times (m-2r)$ identity matrix, and $\Gamma$ is called cosine-sine matrix. An $n$-qudit gate corresponds to a $d^n\times d^n$ unitary matrix. The synthesis of qudit quantum circuits based on CSD was first proposed by Khan \emph{et al}. \cite{27,28}. There, they choose the partition size $r=d^{n-1}$ at each recursion level. Different from the Khan \emph{et al}.'s method, we choose the partition size $r=[d/2]d^{n-1}$ for the first level decomposition, then $r=[d/4]d^{n-1}$ for the second level decomposition, and $r=[d/2^k]d^{n-1}$  for the $k$'th level decomposition, here [\emph{a}] denotes the integer part of \emph{a}.  After $\kappa$ levels $(\log_2d\leq\kappa<\log_2d+1)$ of decomposition, $d^{n-1}\times d^{n-1}$ block diagonal matrices are obtained. The block diagonal matrices correspond to uniformly controlled  $(n-1)$-qudit ($u\Lambda_1(U^{n-1})$) gates; the cosine-sine matrices corresponds to uniformly  $(n-1)$-fold controlled  $R_y(u\Lambda_{n-1}(R_y))$ rotations.

 The second phase of multivalued QSD is the further decomposition for the uniformly controlled  $(n-1)$-qudit gate. It can be decomposed into $d$ copies of  $(n-1)$-qudit gates and $d-1$ copies of controlled  $(n-1)$-qudit diagonal $(\Lambda_1(\Delta^{n-1}))$ gates. In qubit case, the uniformly controlled  $(n-1)$-qudit gate is decomposed into a pair of  $(n-1)$-qudit gate and a $\Lambda_1(\Delta^{n-1})$ gate, and it is equivalent to the decomposition of the block diagonal matrix in QSD. So the decomposition given here is a generalization of QSD for qubit case.

The synthesis of a generic $n$-qudit gate involves three kinds of component:  $(n-1)$-qudit gates,  $\Lambda_1(\Delta^{n-1})$ gates,  $u\Lambda_{n-1}(R_y)$ rotations. The  $(n-1)$-qudit gates can be further decomposed in similar ways. So we can construct a generic $n$-qudit quantum circuit by a recursive way. All the component elements required can be efficiently synthesized based on GCX gate.

\subsection{Optimization of the Synthesis Stated Above} \label{sec-3-2}

\emph{Optimizing the decomposition of matrices}: When $d$ is not a power of 2, there are identity submatrices in the cosine-sine matrices of CSD. We can rearrange and change the block diagonal matrices of CSD to reduce the numbers of two components, the  $(n-1)$-qudit gate and the $\Lambda_1(\Delta^{n-1})$ gate. The number of  $(n-1)$-qudit gates can be reduced to the minimum $d^2$.

For example, an $n$-qutrit gate, corresponding to a $3^n\times3^n$ unitary matrix, can be decomposed as follows
\begin{eqnarray}   \label{eq-5}
W_1=A\Gamma_1B\Gamma_0C\Gamma_2D.
\end{eqnarray}
with
\begin{eqnarray}    \label{eq-6}
&&\Gamma_0=\left(\begin{array}{ccc}
C   &  -S & 0\\
S   &  C  & 0\\
0   &  0  & I\\
\end{array}\right),\quad\quad
\Gamma_1=\left(\begin{array}{ccc}
I   &  0    &  0\\
0   &  C_1  & -S_1\\
0   &  S_1  & C_1\\
\end{array}\right),\nonumber\\
&&\Gamma_2=\left(\begin{array}{ccc}
I   &  0    &  0\\
0   &  C_2  & -S_2\\
0   &  S_2  & C_2\\
\end{array}\right),
\end{eqnarray}
\begin{eqnarray}     \label{eq-7}
&&A=\left(\begin{array}{ccc}
U_1 &  0    & 0\\
0   &  U_2  & 0\\
0   &  0    & U_3\\
\end{array}\right),\quad
B=\left(\begin{array}{ccc}
I &  0    & 0\\
0   &  X_2  & 0\\
0   &  0    & X_3\\
\end{array}\right),\nonumber\\
&&C=\left(\begin{array}{ccc}
V_1 &  0    & 0\\
0   &  V_2  & 0\\
0   &  0    & V_3\\
\end{array}\right),\quad
D=\left(\begin{array}{ccc}
I &  0    & 0\\
0   &  Y_2  & 0\\
0   &  0    & Y_3\\
\end{array}\right),
\end{eqnarray}
where each block matrix in the decomposition above is of size $3^{n-1}\times3^{n-1}$. We can rewrite it as that
\begin{eqnarray}   \label{eq-8}
W_1=A'\Gamma_1B'\Gamma_0C'\Gamma_2D'.
\end{eqnarray}
with
\begin{eqnarray}     \label{eq-9}
&&A'=(I\otimes U_2)\left(\begin{array}{ccc}
I &  0    & 0\\
0   &  I  & 0\\
0   &  0    & U'_3\\
\end{array}\right), \;
B'=(I\otimes X_2)\left(\begin{array}{ccc}
X_1 &  0    & 0\\
0   &  I  & 0\\
0   &  0    & I\\
\end{array}\right),\nonumber\\
&&C'=(I\otimes V_2)\left(\begin{array}{ccc}
I &  0    & 0\\
0   &  I  & 0\\
0   &  0    & V'_3\\
\end{array}\right), \;\;
D'=\left(\begin{array}{ccc}
Y_1 &  0    & 0\\
0   &  Y_2  & 0\\
0   &  0    & Y_3\\
\end{array}\right).
\end{eqnarray}
where $U'_3=U_2^{-1}U_3$,  $X_1=X_2^{-1}U_2^{-1}U_1$,  $V'_3=V_2^{-1}X_2^{-1}X_3V_3$,  $Y_1=V_2^{-1}X_1$. For matrices $A$, $B$, $C$, $D$ and $D'$, each of them corresponds to two $\Lambda_1(\Delta^{n-1})$  gates and three  $(n-1)$-qutrit gates; for $A'$, $B'$, $C'$ each matrix corresponds to one $\Lambda_1(\Delta^{n-1})$ gate and two  $(n-1)$-qutrit gates. The optimal synthesis of a generic $n$-qutrit circuit gate involves nine  $(n-1)$-qutrit gates and five $\Lambda_1(\Delta^{n-1})$ gates, three  $(n-1)$-qutrit gates and three $\Lambda(\Delta^{n-1})$ gates less than that in the original synthesis.

For example again, taking $d=6$, the $n$-qudit circuit, corresponding to a $6^n\times 6^n$ unitary matrix, can be decomposed as follows
\begin{eqnarray}     \label{eq-10}
W_2=A \Gamma_1 B \Gamma_2 C \Gamma_3 D \Gamma_0 E \Gamma_4 F \Gamma_5 G \Gamma_6 H
\end{eqnarray}
with
\begin{eqnarray}     \label{eq-11}
\Gamma_0=\left(\begin{array}{cc}
C   &  -S   \\
S   &  C  \\
\end{array}\right),
\end{eqnarray}
where each bock matrix in Eq. (\ref{eq-11}) is of size  $\frac{6^n}{2} \times \frac{6^n}{2}$, and
\begin{eqnarray}     \label{eq-12}
\Gamma_1=\left(\begin{array}{cccccc}
I   &  0    &  0     &  0  &  0     &  0 \\
0   &  C_1  &  -S_1  &  0  &  0     &  0 \\
0   &  S_1  &  C_1   &  0  &  0     &  0 \\
0   &  0    &  0     &  I  &  0     &  0 \\
0   &  0    &  0     &  0  &  C_2   &  -S_2 \\
0   &  0    &  0     &  0  &  S_2   &  C_2 \\
\end{array}\right), \nonumber\\
\Gamma_2=\left(\begin{array}{cccccc}
C'_1  &  -S'_1  &  0  &  0    &  0     &  0 \\
S'_1  &  C'_1   &  0  &  0    &  0     &  0 \\
0    &  0     &  I  &  0    &  0     &  0 \\
0    &  0     &  0  &  C'_2  &  -S'_2  &  0 \\
0    &  0     &  0  &  S'_2  &  C'_2   &  0 \\
0    &  0     &  0  &  0    &  0     &  I \\
\end{array}\right), \nonumber\\
\Gamma_3=\left(\begin{array}{cccccc}
I   &  0    &  0     &  0  &  0     &  0 \\
0   &  C''_1  &  -S''_1  &  0  &  0     &  0 \\
0   &  S''_1  &  C''_1   &  0  &  0     &  0 \\
0   &  0    &  0     &  I  &  0     &  0 \\
0   &  0    &  0     &  0  &  C''_2   &  -S''_2 \\
0   &  0    &  0     &  0  &  S''_2   &  C''_2 \\
\end{array}\right),
\end{eqnarray}
\begin{eqnarray}     \label{eq-13}
&&A=diag\{A_1,\; A_2,\; A_3,\; A_4,\; A_5,\; A_6\}, \nonumber\\
&&B=diag\{I,\; B_2,\; B_3,\; I,\; B_5,\; B_6\},  \nonumber\\
&&C=diag\{C_1\; C_2,\; C_3,\; C_4,\; C_5,\; C_6\},\nonumber\\
&&D=diag\{I,\; D_2,\; D_3,\; I,\; D_5,\; D_6\},
\end{eqnarray}
where each block matrix in Eqs.(\ref{eq-12}-\ref{eq-13}) is of size  $6^{n-1}\times6^{n-1}$. The second half of expression in Eq. (\ref{eq-10}), $E \Gamma_4 F \Gamma_5 G \Gamma_6 H$, has same form as the first half, $A \Gamma_1 B \Gamma_2 C \Gamma_3 D$. The first half can be rewritten as
\begin{eqnarray}     \label{eq-14}
A \Gamma_1 B \Gamma_2 C \Gamma_3 D  =A' \Gamma_1 B' \Gamma_2 C' \Gamma_3 D'
\end{eqnarray}
with
\begin{eqnarray}     \label{eq-15}
&&A'=(I\otimes A_2)diag\{I, I, A'_3, I, A'_5, A'_6\},\nonumber\\
&&B'=(I\otimes B_2)diag\{B_1', I, I, B'_4, B'_5, I\},\nonumber\\
&&C'=(I\otimes C_2)diag\{I, I, C'_3, I, C'_5, C'_6\},\nonumber\\
&&D'=diag\{D'_1, D_2, D_3, D'_4, D_5, D_6\}.
\end{eqnarray}
Here
$A'_3=A_2^{-1}A_3$,
$A'_5=A_2^{-1}A_5$,
$A'_6=A_2^{-1}A_6$,
$B'_1=B_2^{-1}A_2^{-1}A_1$,
$B'_4=B_2^{-1}A_2^{-1}A_4$,
$B'_5=B_2^{-1}B_5$,
$C'_3=C_2^{-1}B_2^{-1}B_3C_3$,
$C'_5=C_2^{-1}C_5$,
$C'_6=C_2^{-1}B_2^{-1}B_6C_6$,
$D'_1=C_2^{-1}C_1$
and
$D'_4=C_2^{-1}C_4$.
The second half of the expression can be processed in same way. The optimal synthesis of a generic $n$-qudit circuit gate with $d=6$ involves 36  $(n-1)$-qudit gates and 28  $\Lambda_1(\Delta^{n-1})$ gates, eight  $(n-1)$-qudit gates and eight $\Lambda_1(\Delta^{n-1})$ gates less than that in the original synthesis.

\begin{table}[htb]
\centering \caption{Numbers of three components for optimal $n$-qudit quantum circuits.}

\begin{tabular}{lcccccc}
\hline  \hline

           & \multicolumn {6}{c}{$d$} \\
\cline{2-7}
   component               &    3     &     4     &     5     &     6      &    7      &     8        \\
\hline

 $(n-1)$-qudit gate        &    9     &    16      &    25     &      36     &    49     &       64     \\

$\Lambda_1(\Delta^{n-1})$  &    5    &     12     &     17     &      28     &    41    &        56     \\

$u\Lambda_{n-1}(R_y)$    &    3    &     6      &     10    &       15    &     21    &        28    \\
                             \hline  \hline
\end{tabular}\label{Table-1}
\end{table}

The numbers of three components needed to construct optimal generic multivalued quantum circuits are listed in Tab. \ref{Table-1}. A cosine-sine matrix for a qudit system can involve several sets of $u\Lambda_k(R_y)$ rotation. To reduce the number of  $(n-1)$-qudit gates to its minimum $d^2$ is essential for the asymptotic optimality of the synthesis. The number of the GCX gate in these $d^2$  $(n-1)$-qudit gate components account for vast majority of the GCX gate count of an $n$-qudit gate if $n$ is large. For example£¬the number of the GCX gate in nine 2-qutrit gate components account for 68.02\% of the GCX gate count of a generic 3-qutrit circuit. Whereas the number of GCX gate in nine 7-qutrit gate components account for 99.87\% of the count of a generic 8-qutrit circuit. The numbers of two other components may be neglected if $n$ is enough large. Hence the synthesis obtained here is asymptotically optimal, which means that the generic $n$-qudit circuit can be synthesized asymptotically by $O(\alpha d^{2n})$  two-qudit elementary gates, here $\alpha$ is a constant.

\begin{figure}[!h]
\begin{center}
\includegraphics[width=8.0 cm,angle=0]{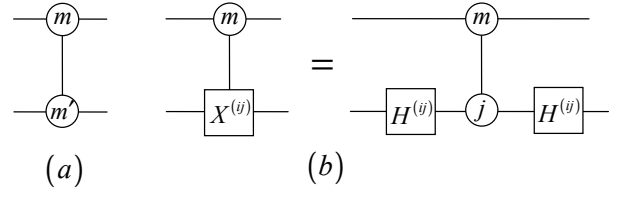}
\caption{GCZ gate (a) and transformation between GCX gate and GCZ gate (b).}
\label{Fig-2}
\end{center}
\end{figure}

\emph{Optimizing the uniformly multifold controlled $R_y$ rotations}: The optimization of the qubit $u\Lambda_{n-1}(R_y)$ rotations has been given in Ref. \cite{19} by using controlled-$Z$ (CZ) gates. To optimize the qudit $u\Lambda_{n-1}(R_y)$  rotations needs the high dimensional counterpart of CZ gate. The multivalued extension of $Z$ operation is a one-qudit operation $Z^{[m]}$ which is specified by that $Z^{[m]}=\sum_{k\neq m}|k\rangle\langle k|-|m\rangle\langle m|$. There are only $d$ different forms of $Z^{[m]}$  operation for a qudit ($d=0,1,\cdots, d-1$, respectively), whereas  there are $d(d-1)/2$ forms of $X^{(ij)}$  operation for the qudit. Like the GCX gate, the generalized controlled $Z$ (GCZ) gate (denoted as $\text{GCZ}(m-m')$) is defined as a controlled 2-qudit gate which implements the $Z^{[m']}$  operation on the target qudit iff the control qudit is in the state $|m\rangle$. It is specified by  $\text{GCZ}(m-m')=\sum_{ij\neq mm'}|ij\rangle\langle ij|-|mm'\rangle\langle mm'|$. Like the qubit case, the control qudit and target qudit of the GCZ gate are changeable, and using two generalized Hadamard gates, the GCX and GCZ gates can be transformed each other.  The generalized Hadamard gate $H^{(ij)}$  is a one-qudit gate specified by that  $H^{(ij)}=\sum_{k\neq i,j}|k\rangle\langle k|+(|i\rangle\langle i|+|i\rangle\langle j|+|j\rangle\langle i|-|j\rangle\langle j|)/\sqrt{2}$. The circuit representation of GCZ gate and its transformation relation with GCX gate are shown in Fig. \ref{Fig-2}.
%

 The statements and Fig. 14 in Appendix C of Ref. \cite{31} still hold for qudit $u\Lambda_{n-1}(R_y)$ rotations if all  $\text{GCX}(m\rightarrow X^{(ij)})$ gates are replaced with $\text{GCZ}(m-j)$ gates. Thus a set of qudit $u\Lambda_{n-1}(R_y)$ rotation may be implemented with $2d^{n-2}(d-1)$  GCZ gates, of which   $d-1$ $\text{GCZ}(m-j)$  gates ($m=1,2,\cdots,d-1$, respectively) may be moved furthest to the right (or the left). The rightmost   $d-1$ GCZ gates produce a diagonal gate which may be absorb into the neighboring uniformly controlled  $(n-1)$-qudit gate. The reason why the $d-1$  GCZ gates are able to be moved furthest to the right whereas only one CZ gate is able to move like that in qubit case is that the controlled gates with different control basis states can exchange one another. This saves $d-1$  two-qudit elementary gates  for each set of qudit $u\Lambda_k(R_y)$ rotation, totally save $(d^{2(n-1)}-1)n_{u R}/(d+1)$  GCX gates for a generic $n$-qudit circuit, where  $n_{u R}$ is the number of $u\Lambda_{n-1}(R_y)$  component in the circuit and is given in Tab. \ref{Table-1}. In the practical process of optimization, it should optimize $u\Lambda_{n-1}(R_y)$ rotations first, then to optimize the decomposition of matrices.

\subsection{Structure and GCX Gate Count of Optimal Circuits} \label{sec-3-3}

\begin{figure*}[htp]
\begin{center}
\includegraphics[width=13.5 cm,angle=0]{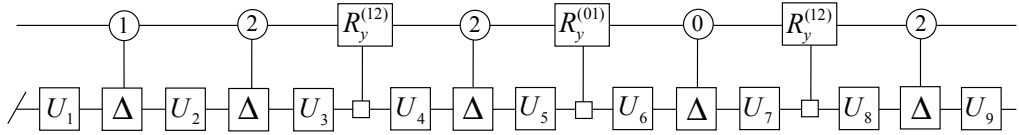}
\caption{Structure of a generic $n$-qutrit circuit. Here the small square ($\boxempty $)denotes the uniform control, the slash (/) represents multiple qutrits on the line.}
 \label{Fig-3}
\end{center}
\end{figure*}

The optimal quantum circuit of generic $n$-qudit circuits involves $d^2$  $(n-1)$-qudit gates, which are separated by $\Lambda_{1}(\Delta^{n-1})$ gates or circuits for cosine-sine matrix. It involves $d^2-2^\kappa$ $\Lambda_{1}(\Delta^{n-1})$   gates and $2^\kappa-1$  circuits for cosine-sine matrix. In multivalued case, a circuit for cosine-sine matrix usually involves several sets of $u\Lambda_{n-1}(R_y)$ rotations. The structure of a generic $n$-qutrit circuit is illustrated in Fig. \ref{Fig-3}.

The GCX gate count of the optimal multivalued quantum circuits given here is tabulated in Tab. \ref{Table-2}. For comparison the count before the optimization is also given. The results are obviously improved after optimization, especially for the case $d$ is not a power of two. The circuits optimized have asymptotic optimal features for all dimensionality $d$, whereas the circuits before optimization are not asymptotically optimal except for $d$ is a power of two.

\begin{table*}[htb]
\centering \caption{The GCX gate count for the synthesis of qudit quantum circuits obtained using the multivalued QSD. In each cell, the upper line denotes the count before optimization \cite{31}, the bottom line denotes the count optimized.}

\begin{tabular}{lcccccc}
\hline  \hline

           & \multicolumn {6}{c}{$d$} \\
\cline{2-7}
  $n$              &    3     &     4     &     5     &     6      &    7      &     8        \\
\hline

2  &   \tabincell{l}{44\\ 26}   &  \tabincell{l}{108\\90}   &  \tabincell{l}{272\\176}   &  \tabincell{l}{510\\355}   &   \tabincell{l}{828\\618}   &    \tabincell{l}{1176\\980} \\

3  &   \tabincell{l}{692\\344}   &  \tabincell{l}{2232\\1926}   &  \tabincell{l}{10256\\5216}   &  \tabincell{l}{25860\\15565}   &   \tabincell{l}{52740\\53856}   &    \tabincell{l}{85456\\72716}\\

4  &   \tabincell{l}{6860\\3458}   &  \tabincell{l}{37800\\32886}   &  \tabincell{l}{336144\\136576}   &  \tabincell{l}{1158720\\577705}   &   \tabincell{l}{2965788\\1797210}   &    \tabincell{l}{5551504\\4735948} \\

5  &   \tabincell{l}{83924\\32028}   &  \tabincell{l}{613248\\534582}  & \tabincell{l}{10796560\\3445576}  & \tabincell{l}{51109320\\20902225}  &  \tabincell{l}{166400964\\88346400}  &  \tabincell{l}{355955600\\303759820}  \\

6  &   \tabincell{l}{1011932\\291638} & \tabincell{l}{9854970\\8587062} & \tabincell{l}{345689872\\86295576} & \tabincell{l}{$2.25\times10^9$\\752674425} & \tabincell{l}{$9.32\times10^9$\\$4.33\times10^9$} &  \tabincell{l}{$2.28 \times10^{10}$\\$1.95\times10^{10}$}  \\

7  &   \tabincell{l}{12157748\\2634932} & \tabincell{l}{$1.58\times10^8$\\$1.35\times10^8$} & \tabincell{l}{$1.11\times10^{10}$\\$2.16\times10^9$} & \tabincell{l}{$9.90\times10^{10}$\\$2.71\times10^{10}$} & \tabincell{l}{$5.22\times10^{11}$\\$2.12\times10^{11}$} & \tabincell{l}{$1.46\times10^{12}$\\$1.25\times10^{12}$}    \\

8  &   \tabincell{l}{$1.46\times10^8$\\23744984}   &  \tabincell{l}{$2.52\times10^9$\\$2.20\times10^9$}   &  \tabincell{l}{$3.55\times10^{11}$\\$5.39\times10^{10}$}   &  \tabincell{l}{$4.36\times10^{12}$\\$9.75\times10^{11}$}   &   \tabincell{l}{$2.92\times10^{13}$\\$1.04\times10^{13}$}   &    \tabincell{l}{$9.34\times10^{13}$\\$8.00\times10^{13}$} \\

                             \hline  \hline
\end{tabular}\label{Table-2}
\end{table*}

\section{ Efficiency of Synthesis of Quantum Circuits}   \label{sec-4}

To evaluate the efficiency of synthesis of generic $n$-qudit circuits based on GCX gates, we propose a quantity known as efficiency index ($\mathscr{R}$) which is defined by  $\mathscr{R}=d^{2n}/N_n$, where $N_n$ is the number of GCX gate required to synthesize the $n$-qudit circuit. The quantity   $\mathscr{R}$ is the average number of parameters carried by each GCX gate. The larger the quantity  $\mathscr{R}$, the more efficient the synthesis of the circuit is. For the quantum circuit synthesis which is asymptotically optimal, there is an asymptotic efficiency index  $\mathscr{R}^{\rm{asy}}$  which is the efficiency index when n is enough large. The asymptotic efficiency indexes for the optimal synthesis given here are listed in Table III. From the Table, it can be seen that the $\mathscr{R}^{\rm{asy}}$  is increased with the increase of dimensionality $d$.

\begin{table}[htb]
\centering \caption{The asymptotic efficiency indexes for the optimal synthesis of multivalued quantum circuits.}

\begin{tabular}{lcccccc}
\hline  \hline
  $d$                     &    3     &     4     &     5     &     6      &    7      &     8        \\
\hline

 $\mathscr{R}^{\rm{asy}}$ &    1.81     &    1.95     &    2.48    &     2.89     &    3.19     &       3.53     \\

                             \hline  \hline
\end{tabular}\label{Table-3}
\end{table}

There are several previous works on the synthesis of multivalued quantum circuits based on elementary gates. Based on the controlled increment (CINC) gate, the synthesis by using the spectrum decomposition algorithm is investigated in Refs. \cite{23,25}. It is asymptotic optimal, which has leading factor 2 for the CINC account of the synthesis. Using the GCX gate as the two-qudt elementary gate instead of the CINC gate, the synthesis is greatly simplified \cite{31}. The synthesis simplified still has leading factor 2 but for the GCX account. So it's $\mathscr{R}^{\rm{asy}}$  is equal to 0.5 for all dimensionality $d$, which less than all values of $\mathscr{R}^{\rm{asy}}$ in Tab. \ref{Table-3}.

\begin{table*}[htb]
\centering \caption{The CINC gate count for the synthesis of qudit quantum circuits obtained by using CSD with balanced partition \cite{28}.}

\begin{tabular}{lcccccc}
\hline  \hline

           & \multicolumn {6}{c}{$d$} \\
\cline{2-7}
  $n$              &    3     &     4     &     5     &     6      &    7      &     8        \\
\hline

2  &  36  & 72  & 280  & 420  &  588  &   784 \\

3  & 3360  &  4464  &  69720 & 60960  & 381780  & 142240  \\

4  & 20088 & 40824 & 670320 & 1563660  & 4928616  & 4750256  \\

5  & 1382952 & 685440 & 252347440 & 38074200 & $1.0\times10^{10}$  & $3.1\times10^8$  \\

6  & 8254764 & 22254984  & $2.5\times10^9$  & $3.7\times10^9$ & $1.4\times10^{11}$  & $3.9\times10^{10}$   \\

7  & 127837404 & 357389712  & $1.4\times10^{11}$  & $1.8\times10^{11}$  & $1.1\times10^{13}$  & $2.5\times10^{12}$  \\

8  & 465572880 & $2.9\times10^9$  & $8.8\times10^{11}$  & $4.2\times10^{12}$  & $9.8\times10^{13}$  & $8.0\times10^{13}$  \\

                             \hline  \hline
\end{tabular}\label{Table-4}
\end{table*}

Based on CINC gate, the syntheses by using the CSD with balanced partition are investigated in Ref. \cite{28}. The circuits synthesized by this method are simpler than those by using the spectrum decomposition if $n$ is small, but they are not asymptotically optimal except for $d$ is a power of two. The results of this work are given in Tab. \ref{Table-4}. It needs $d-1$  GCX gates to synthesize a CINC gate \cite{31}. Comparing the data in Tab. \ref{Table-2} and that in Tab. \ref{Table-4}, considering the CINC gate itself has complex construct, it can be seen that the synthesis of quantum circuits given in this article are much more efficient than that in Ref. \cite{28} even if the $n$ is very small.

\begin{table}[htb]
\centering \caption{The CDNOT gate count for the synthesis of ququart quantum circuits \cite{29}.}

\begin{tabular}{lcccccc}
\hline  \hline
  $n$                     &    2     &     3     &     4     &     5      &    6      &     $n$        \\
\hline

 Gate count &    60     &   1200    &   20160    &   326400     &   $5.2\times10^6$    &  $(5/16)4^{2n}\!-\!(5/4)4^n$     \\

                             \hline  \hline
\end{tabular}\label{Table-5}
\end{table}

 Li \emph{et al}. propose a two-ququart gate, termed the controlled double-not (CDNOT) gate, for four level quantum systems. Based on the CDNOT gate, they investigate the synthesis of ququart quantum circuits by using QSD method \cite{29}, the results are tabulated in Tab. \ref{Table-5}. A CDNOT gate is a two-ququart controlled gate which implements the $\sigma_x\otimes I_2$  operation on the target ququart iff the control ququart is in the state $|m\rangle$,  $m\in2,3$, here $\sigma_x$  is a Pauli matrix. The  $\sigma_x\otimes I_2$ operation is equivalent to two $X$ operations: $X^{(02)}$ and  $X^{(13)}$, so a CDNOT gate is equivalent to two GCX gate. The $\mathscr{R}^{\rm{asy}}$ of this synthesis is 1.60, still less than that in Tab. \ref{Table-3} for ququart 1.95. From discussion above, it can be seen that our algorithm given here is first efficient multivalued synthesis routine which produces best known results for all dimensionality $d$, and for both the small $n$ case and the asymptotic case.

The syntheses of generic $n$-qubit circuits based on QSD and their asymptotic efficiency indexes are listed in Tab. \ref{Table-6}. 
The qubit counterpart of the optimal synthesis for the generic $n$-qudit circuits given here is the qubit circuits based on QSD with recursion bottom out at the one-qubit circuit ($l=1$) and the optimization for $u\Lambda_{n-1}(R_y)$ rotations (the second line of Tab. \ref{Table-6}), its $\mathscr{R}^{\rm{asy}}$ is 1.50, less than all asymptotic efficiency indexes in Tab. \ref{Table-3}. For the qubit case, there is a most efficient synthesis for generic two-qubit circuits which reaches its theoretical lower bound of complexity three CNOT gates. The best result for the synthesis of generic $n$-qubit circuits is based on QSD with recursion bottom out at the two-qubit circuit ($l=2$) and two additional optimizations (the fourth line of Tab. \ref{Table-6}), its $\mathscr{R}^{\rm{asy}}$ is 2.09. Now the asymptotic efficiency indexes of the generic $n$-qudit circuits with  $d\geq5$  have been greater than this value. Moreover the generic two-ququart circuit has been more efficient than the most efficient generic four-qubit circuit.

\begin{table}[htbp]
\caption {The CNOT counts of $n$-qubit quantum circuits based on QSD.}

\begin{tabular}{lccccclc}

\hline\hline
                & 2      & 3      & 4       & 5      & 6    & $n$                                        &   $\mathscr{R}^{\rm{asy}}$     \\

$l\!\!=\!\!1$\, \cite{19} & 6     & 36     & 168      &720     & 2976 & $(3/4)\!\times\! 4^{2n}\!\!-\!(3/2)\!\times\! 4^n\! $      &  1.33 \\

$l\!\!=\!\!1,\!\text{\footnotesize{optimal}} $ & 5     & 31     & 147      & 635    & 2635 & $(2/3)\!\times\! 4^{2n}\!-\!(3/2)\!\times\! 4^n\! +\!1/3\!$   & 1.50 \\

$l\!\!=\!\!2$ \cite{19} & 3     & 24     & 120      & 528    & 2208 & $(9/16)\!\times\! 4^{2n}\!-\!(3/2)\!\times\! 4^n\!$       & 1.78 \\

$l\!\!=\!\!2,\!\text{\footnotesize{optimal}}$\cite{19}  & 3     & 20     & 100       & 444   & 1868 & $(23/48)\!\times\!\! 4^{2n}\!\!-\!\!(3/2)\!\!\times\! 4^n\!\! +\!4 /3\!\! $ & 2.09 \\

\hline\hline

\end{tabular}               \label{Table-6}
\end{table}

\section{Conclusion and future work}   \label{sec-5}

We have optimized the synthesis of generic multivalued quantum circuits. The optimal circuits are asymptotically optimal for all dimensionality $d$, so that we can build efficient quantum circuits for the qudit $d$ is not a power of two as that $d$ is a power of two. It is of great significance to make full use of various resources. The algorithm given here is the most efficient qudit synthesis routine so far which produces best known results in all respects.

The multivalued quantum circuits do have advantages over the binary quantum circuits. The generic $n$-qudit circuits with $d\geq5$  and generic two-ququart circuits given here are practical circuits which are more efficient than the most efficient qubit circuits. The leading factor of the lower bound of complexity for qudit circuits is small by a factor of $d-1$ in comparison to that for qubit circuits and the asymptotic efficiency index is increased with the increase of dimensionality $d$, further reveal the advantages and benefits of qudit circuits over generic qubit circuits.

There is still plenty of room for improvement in the synthesis of multivalued quantum circuits. One of most important work for the improvement is to optimize the two-qudit quantum circuits. Since our algorithm for generic qudit circuits given here is recursive, the more efficient generic qudit circuits can be obtained from more efficient two-qudit circuits.

\section*{ACKNOWLEDGEMENTS}

This work is supported by the National Natural Science Foundation of China under Grant No. 11204112 \& 11447015 and the Priority Academic Program for the Development of Jiangsu Higher Education Institutions.


\end{document}